\documentclass[11pt,a4paper]{article}
\usepackage{amscd,amsmath,amssymb,latexsym,bbm}
\usepackage{amssymb}
\usepackage{graphicx}

\begin{document}

\title{A rigorous proof of non-existence of edge state in the semi-infinite
armchair edged graphene}
\author{Yuanyuan Zhao, Wei Li and Ruibao Tao \\
State Key Laboratory of Surface Physics and Department
of Physics,\\ Fudan University, Shanghai, 200433, People's Republic of China}
\maketitle

\begin{abstract}
With the help of transfer matrix method, the conditions for the existence of 
the edge states in the semi-infinite armchair edged graphene is given. We discuss 
zero-energy and non-zero-energy edge states separately, and show the nonexistence 
of the edge states in the model analytically and rigorously.
\end{abstract}


\section{Introduction}

One of the interesting phenomena in solid state physics is the
existence of edge states on the boundary, the properties of which
are distinct from those of the bulk states, and they can play
an important role in transport, there are examples showing that
system is insulated in the bulk, while conduction can be allowed by
edge states on the boundary. The most prominent ones are the quantum
Hall effect (QHE)\cite{QHE, LaughlinQHE, HalperinQHE, Bhatt} and the
quantum spin Hall effect (QSHE)\cite{Kane1, Bernevig1, Bernevig2},
where the quantization of a Hall conductance is tightly associated
with the edge states\cite{LaughlinQHE, HalperinQHE, Bhatt, Kane1,
Bernevig1, Bernevig2, Thouless, Kohmoto, Hatsugai1}.

The realization of graphene\cite{Novoselov1, YZhang, Geim}
in laboratories has triggered great research interests in recent years.
From the topological viewpoint, edge states can be induced in the system
with different edge geometries\cite{Klein, Fujita}; in experiments,
with the help of scanning tunneling microscopy and scanning tunneling
spectroscopy, the presence of structure-dependent edge states of graphite
can be observed\cite{Niimi,Kobayashi}. There are some
papers\cite{Hatsugai2, Wakabayashi} showing the similarity between
the graphene model and the $d$-wave superconductor\cite{Hu} with
edges, where the existence of edge states depends on its edge shapes.
For comparison, on a $\{110\}$ surface of a $d_{x^{2}-y^{2}}$
-wave superconductor, the origin of the Andreev bound states is the $\pi $
-phase shift due to the unconventional pairing symmetry, and graphene can be
seem as an odd-parity superconductor\cite{Wakabayashi}, at the zigzag/bearded
edges, different signs of pairing potential lead to a $\pi $-phase
shift and zero-energy bound states, while at the armchair edge, it will not
show the localized edge states. Hence the study on the zero mode of the edge
state will have the fundamental meaning that can supply some information
about the phase structure.

The purpose of this paper is to study the edge states of the semi-infinite
armchair edged graphene (AEG) analytically with the help of transfer matrix
method\cite{DHLee, LJiang, Haidong, Li-Tao, YZhao}. There are many papers
showing the edge states in graphene models in different ways, while most of
them focus on the zero mode of the zigzag graphene nanoribbon. And here we
pay our attention on the semi-infinite armchair edged graphene, and study
the zero-energy and non-zero-energy edge states. According to the discussions
of the properties of the transfer matrix, we show an analytically proof of the
non-existence of the edge states in the semi-infinite AEG. We have to mention
that we just consider about nearest-neighbor-interaction in our whole discussions.

\section{ EDGE STATES OF SEMI-INFINITE AEG}

\begin{figure}[h]
\centering\includegraphics[width = 0.6\textwidth]{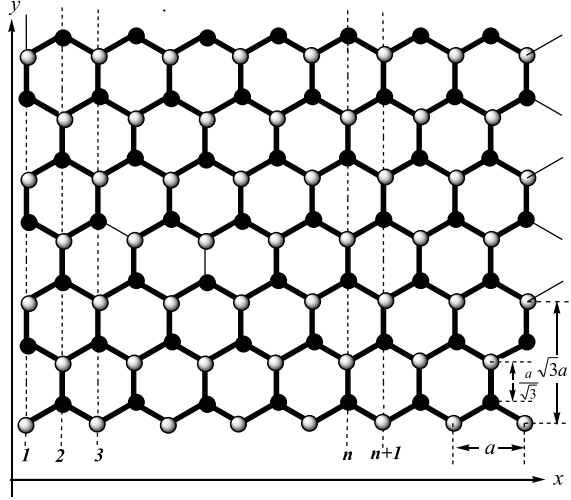}
\caption{Schematic
illustration of the lattice structure of semi-infinite armchair edged graphene. The black circles
represent the type $B$ atom, while empty circle represents type
$A$ atom; It is infinite in $y$ direction and the edge locates at
the left perpendicular chain labeled by number 1.} \label{AEG}
\end{figure}

The geometrical structure of graphene is shown in Fig.\ref{AEG}. It
is infinite in $y $ direction with periodic constant
$\sqrt{3}a$, here $a$ is the lattice constant between two n.n. A(B)
atoms. We arrange all vertical chains in the order from the
left to the right as $\{1,2,3,\cdots \}$. Each vertical chain is an 1D
atomic chain, in which the periodic cell contains two kinds of atoms $A$ and
$B.$ The position of the atoms are labeled by two indices $(n,j)$,
where $n$ labels the order number of vertical chains from $1$ to
$\infty$, with $n=1$ representing the left edge.

Due to its infinity in $y$ direction, $k_{y}$ is a good quantum number,
so that we can take the Fourier transformation on wave functions $\Psi
_{A(B)}^{(n,j)}(k_{y})$ for $A(B)$ atoms: $\Psi
_{A(B)}^{(n,j)}(k_{y})=\exp (ik_{y}y_{j,A(B)})\Phi
_{n,A(B)}(k_{y})$. The Hamiltonian now in $\{n,k_{y}\}$ representation can
be expressed by a set of Fermion operators $\{\Phi_{n,A(B)}(k_{y})\},n=1,2,\ldots
,\infty $:

\begin{eqnarray}  \label{eq:2-1}
H &=&t\sum_{k_{y},n\geqslant 1}e^{ik_{y}a/\sqrt{3}}\Phi_{n,A}^{\dag
}(k_{y})\Phi_{n,B}(k_{y})  \notag \\
&+&t\sum_{k_{y},n\geqslant 2}e^{-ik_{y}a/2\sqrt{3}}\Phi_{n,A}^{\dag
}(k_{y})\Phi_{n-1,B}(k_{y})  \notag \\
&+&t\sum_{k_{y},n\geqslant 1}e^{-ik_{y}a/2\sqrt{3}}\Phi_{n,A}^{\dag
}(k_{y})\Phi_{n+1,B}(k_{y}) +~h.c.
\end{eqnarray}

From the dynamical equations of $\{\Phi_{n,A(B)}(k_{y})\}$:
$E\Phi_{n,A(B)}(k_{y})=[\Phi_{n,A(B)}(k_{y}),H],$ we get
following recursive equations for each fixed wave vector $k_{y}$:
\begin{eqnarray}  \label{eq:2-2}
\begin{cases}
E\Phi_{n,A} &=\quad te^{ik_{y}a/\sqrt{3}}\Phi_{n,B}+te^{-ik_{y}a/2\sqrt{
3}}\Phi_{n-1,B} \\[0.1cm]
& + te^{-ik_{y}a/2\sqrt{3}}\Phi_{n+1,B},~n = 2,3,\cdots, \\
E\Phi_{n,B} &=\quad te^{-ik_{y}a/\sqrt{3}}\Phi_{n,A}+te^{ik_{y}a/2\sqrt{
3}}\Phi_{n-1,A} \\
& + te^{ik_{y}a/2\sqrt{3}}\Phi_{n+1,A},~n = 2,3,\cdots;
\end{cases}
\end{eqnarray}
where we use $\Phi_{n,A(B)}$ instead of $\Phi_{n,A(B)}(k_y)$. The equations of the wave functions for the boundary
sites $(n=1)$ are
\begin{eqnarray}  \label{eq:2-3}
\begin{cases}
E\Phi_{1,A} &=\quad te^{ik_{y}a/\sqrt{3}}\Phi_{1,B}+te^{-ik_{y}a/2
\sqrt{3}}\Phi_{2,B}, \\
E\Phi_{1,B} &=\quad te^{-ik_{y}a/\sqrt{3}}\Phi_{1,A}+te^{ik_{y}a/2
\sqrt{3}}\Phi_{2,A}.
\end{cases}
\end{eqnarray}
We can get the bulk energy for the extended states, according
to Eq. (\ref{eq:2-2}), take $k_{x}$ as good quantum number,
the bulk energy satisfies:
\begin{equation}\label{eq:bulk}
    E^{2} = 1+4\cos^{2}{(k_{x}a/2)}+4\cos{(\sqrt{3}k_{y}a/2)}\cos{(k_{x}a/2)}
\end{equation}
which is the energy dispersion relation of the 2D graphene system. In the following,
we will focus on the edge states, and discuss the zero-energy ($E=0$)
and non-zero-energy ($E\neq 0$) cases, separately. To examine whether the edge states
can exist in the semi-infinite graphene model or not.

(1) $E=0$: the zero-energy spectrum. The recursions Eq. (\ref{eq:2-2})
are reduced into two decoupled equations for sublattice $A$
and $B$:
\begin{eqnarray}  \label{eq:2-4}
\begin{cases}
\Phi_{n+1,\alpha} &=\quad -e^{-i\eta_{\alpha}\sqrt{3}k_{y}a/2}\Phi_{n,\alpha}-\Phi_{n-1,\alpha}, \\
\Phi_{2,\alpha} &=\quad -e^{-i\eta_{\alpha}\sqrt{3}k_{y}a/2}\Phi_{1,\alpha}, \\
& \alpha=A,B,~\eta_A=1,~\eta_B=-1,~~n=2,3,\cdots.
\end{cases}
\end{eqnarray}

Above equations can be rewritten as follows
transfer matrix equations:
\begin{eqnarray}
& \left(
\begin{array}{c}
\Phi_{n+2,\alpha} \\
\Phi_{n+1,\alpha}
\end{array}
\right) =T_{\alpha}^{n}\left(
\begin{array}{c}
b_{\alpha} \\
1
\end{array}
\right) \Phi _{1,\alpha},~ n=1,2,3,\cdots.  \label{eq:2-5} \\
\text{where}~~& b_\alpha =-e^{-i\eta_{\alpha}\sqrt{3}k_{y}a/2}
\label{eq:2-6}
\end{eqnarray}
\begin{eqnarray}  \label{eq:2-7}
T_{\alpha} =\left(
\begin{array}{cc}
-e^{-i\eta_{\alpha}\sqrt{3}k_{y}a/2} & -1 \\
1 & 0
\end{array}
\right)
\end{eqnarray}
$\det T_{\alpha}=1.$ It is easy to get two eigenvalues of
$T_{\alpha }(\alpha =A,B)$: $\lambda_{\pm }^{(\alpha )}
=\frac{1}{2}\Lambda_{\alpha }\pm
\frac{1}{2}i\sqrt{4-\Lambda_{\alpha }^{2}}$ where
$\Lambda_\alpha=e^{-i\eta_\alpha k_{y}a\sqrt{3}/2}$.
$|\Lambda_\alpha|=1$ and two eigenvalues $\lambda_{\pm }^{(\alpha
)}$ must be complex and
$\lambda^{\alpha}_{\pm}=(\lambda^{\alpha}_{\mp})^*$. It results
$|\lambda^{\alpha}_{\pm}|=1$, due to
$\lambda^{\alpha}_{+}\cdot\lambda^{\alpha}_{-}=1$. Thus, we can
conclude that the states are always extended in the semi-infinite
AEG, no zero-energy edge state can exist.

(2) $E\neq 0$:\quad according to Eq. (\ref{eq:2-2}), the equations of
two sublattices are coupled with each other. We can also write down a
transfer matrix for $A$ or $B$ type sublattices, and it becomes $4\times 4$.
For more concise description, we define a $4\times 1$ vector $\Psi
_{\alpha }^{(n)}$, its transpose $(\Psi _{\alpha }^{(n)})^{t}$ is
a $1\times 4$ vector as $\left(
\begin{array}{cccc}
\Phi_{n,\alpha} & \Phi_{n-1,\alpha} & \Phi_{n-2,\alpha} &
\Phi_{n-3,\alpha}
\end{array}
\right) $. Then, the Eq. (\ref{eq:2-2}) can be expressed as follows:
\begin{equation}
\begin{cases}
\Psi _{\alpha}^{(n+2)}=T_{0}\Psi _{\alpha}^{(n+1)},n\geqslant 3 \\
T_{0}=\left(
\begin{array}{cccc}
G & \lambda  & G & -1 \\
1 & 0 & 0 & 0 \\
0 & 1 & 0 & 0 \\
0 & 0 & 1 & 0
\end{array}
\right) ,\quad \det{T_{0}}=1,
\end{cases}
\label{eq:2-8}
\end{equation}
where $G=-2\theta $, $\theta =\cos \left( \sqrt{3}k_{y}a/2\right) $, $
\lambda =(E^{2}-3t^{2})/t^{2}=E^{2}-3$ (set $t=1$). We further introduce one
fictitious lattice line at left of edge chain and set them all to be zero: $
\Phi_{\alpha }^{(0)}=0.$ It will not change anything even
introducing the hopping coupling with $1st$ line. Now
Eq. (\ref{eq:2-2}) and Eq. (\ref{eq:2-3}) can be written as
follows:
\begin{eqnarray}
\Psi_{\alpha }^{(n)} &=& T_{0}^{n-3}\Psi _{\alpha }^{(3)}, \quad n\geqslant 4.
\label{eq:2-9} \\
\Psi_{\alpha}^{(3)} &=& G\Psi_{\alpha}^{(2)}+(\lambda +1)\Psi_{\alpha}^{(1)}.  \label{eq:2-10}
\end{eqnarray}
the eigenvalues of transfer matrix $T_{0}$ can be written in the form of
$\{\lambda_{i}\}$:
\begin{eqnarray}\label{eq:2-11}
& \lambda _{1}=\frac{1}{2}\left( -a_{1}-\sqrt{a_{1}^{2}-4}\right)
,\lambda _{2}=\frac{1}{2}\left( -a_{1}+\sqrt{a_{1}^{2}-4}\right)
,\notag\\
& \lambda _{3}=\frac{1}{2}\left(
-a_{2}-\sqrt{a_{2}^{2}-4}\right),\lambda_{4}=\frac{1}{2}\left(
-a_{2}+\sqrt{a_{2}^{2}-4}\right) , \notag\\
&
\end{eqnarray}
where
\begin{equation}\label{eq:2-12}
a_{1}=\theta +\sqrt{m},\, a_{2}=\theta -\sqrt{m},\, m=\lambda
+2+\theta^{2}.
\end{equation}
Now, define a transformation $U$, where $U^{-1}$ exists, to rewrite the transfer
matrix and its relation Eq. (\ref{eq:2-8}) with
$D=U^{-1}T_{0}U=diag \{\lambda_{1},\lambda _{2},
\lambda _{3},\lambda _{4}\}$. In order to get the physically meaningful states,
we have to ensure that all eigenvalues $\{|\lambda_i| \}$ must be finite for
any finite energy $E$.

When $m \neq 0$, $a_{1}^{2} \neq 4$, and $a_{2}^{2} \neq 4$, we can write down
$U$ as follows
\begin{equation}\label{eq:2-15}
U=\left(
\begin{array}{cccc}
\lambda _{1}^{3} & \lambda _{2}^{3} & \lambda _{3}^{3} & \lambda _{4}^{3} \\
\lambda _{1}^{2} & \lambda _{2}^{2} & \lambda _{3}^{2} & \lambda _{4}^{2} \\
\lambda _{1} & \lambda _{2} & \lambda _{3} & \lambda _{4} \\
1 & 1 & 1 & 1
\end{array}
\right)
\end{equation}
with $\det{U}\neq 0$. Its inverse matrix $U^{-1}$ can be found and its elements
of $U^{-1}_{ij}$ are described as follows
\begin{equation}\label{eq:2-16}
\begin{cases}
U_{ij}^{-1}=C_{ij}/u_{i}(a_{1}-a_{2}), \\
u_{1}=a_{1}^{2}-4+a_{1}\sqrt{a_{1}^{2}-4}=-2\lambda_1\sqrt{a_{1}^2-4}, \\
u_{2}=4-a_{1}^{2}+a_{1}\sqrt{a_{1}^{2}-4}=-2\lambda_2\sqrt{a_{1}^2-4}, \\
u_{3}=a_{2}^{2}-4+a_{2}\sqrt{a_{2}^{2}-4}=-2\lambda_3\sqrt{a_{2}^2-4}, \\
u_{4}=a_{2}^{2}-4-a_{2}\sqrt{a_{2}^{2}-4}=2\lambda_4\sqrt{a_{2}^2-4}.
\end{cases}
\end{equation}
where
\begin{equation}\label{eq:2-17}
\begin{cases}
C_{41}=C_{31}=C_{21}=-C_{11}=2, \\
C_{12}=\sqrt{a_{1}^{2}-4}-a_{1}-2a_{2}, \\
C_{22}=\sqrt{a_{1}^{2}-4}+a_{1}+2a_{2} \\
C_{13}=-2-a_{1}a_{2}+a_{2}\sqrt{a_{1}^{2}-4} \\
C_{23}=2+a_{1}a_{2}+a_{2}\sqrt{a_{1}^{2}-4} \\
C_{33}=2+a_{1}a_{2}-a_{1}\sqrt{a_{2}^{2}-4}, \\
C_{43}=2+a_{1}a_{2}+a_{1}\sqrt{a_{2}^{2}-4}, \\
C_{32}=2a_{1}+a_{2}-\sqrt{a_{2}^{2}-4}, \\
C_{42}=2a_{1}+a_{2}+\sqrt{a_{2}^{2}-4}, \\
C_{14}=-a_{1}+\sqrt{a_{1}^{2}-4},~~C_{24}=a_{1}+\sqrt{a_{1}^{2}-4} \\
C_{34}=a_{2}-\sqrt{a_{2}^{2}-4},~~C_{44}=a_{2}+\sqrt{a_{2}^{2}-4}
\end{cases}
\end{equation}
The relation $U^{-1}T_{0}U=D$ was carefully confirmed by $U$ and
$U^{-1}$. After tedious but straightforward algebraic calculations, we obtain
\begin{equation}\label{eq:2-18}
\Psi_{\alpha }^{(n+3)}=UD^{n}U^{-1}\Psi_{\alpha }^{(3)}=U\left(
\begin{array}{c}
W_{1}\lambda _{1}^{n} \\
W_{2}\lambda _{2}^{n} \\
W_{3}\lambda _{3}^{n} \\
W_{4}\lambda _{4}^{n}
\end{array}
\right) ,n\geqslant 1.
\end{equation}
Where
\begin{eqnarray} \label{eq:2-19}
W_{i} &=&\left( U_{i2}^{-1}+GU_{i1}^{-1}\right) \Phi _{2,\alpha
}+\left( U_{i~\!3}^{-1}+\left( \lambda +1\right)
U_{i1}^{-1}\right) \Phi _{1,\alpha
},  \notag  \\
&&i=1,2,3,4.
\end{eqnarray}
Since all elements of matrix $U_{ij}$ are non-zero and
finite, thus the criteria whether the state $\Phi_{n,\alpha}$
is extended or edge state can be determined by those $\{W_{i}\lambda
_{i}^{n}\}$ regardless the front matrix $U$ in Eq.
(\ref{eq:2-18}).

From the definition of $\{\lambda _{i}\}$, we can confirm $\lambda
_{1}\cdot \lambda _{2}=1$ and $\lambda _{3}\cdot \lambda _{4}=1$.
thus $|\lambda _{2}|=|\lambda _{1}|^{-1}$ and $|\lambda
_{3}|=|\lambda _{4}|^{-1}.$  Define $\{\lambda_1,\lambda_2\}$
($\{\lambda_3,\lambda_4\}$) as a pair partner.
$\lambda_{2(1)}$ is the partner of $\lambda_{1(2)}$, while
$\lambda_{4(3)}$ the partner of $\lambda_{3(4)}$. Then we use
$\lambda_{\overline{i}}$ to denote the partner of $\lambda_i$. For
example, $\lambda_2={\lambda}_{\overline{1}}$ and
$\lambda_1={\lambda}_{\overline{2}}$. It is similar to define
the partner of $W_i$ [see Eq. (\ref{eq:2-18}) and Eq. (\ref{eq:2-19})]:
the partner of $W_{1(2)}$ is
$W_{2(1)}=W_{\overline{1}(\overline{2})}$, and $W_{3(4)}$ the
$W_{4(3)}=W_{\overline{3}(\overline{4})}$. If $|\lambda_i|=1$, we
do have $|\lambda_{\overline{i}}|=1$, which corresponds to the extended
states. As we know, the edge states correspond to some $|\lambda_{i}|<1$,
due to the properties of $\lambda_{i}$ and its partner $\lambda_{\overline{i}}$,
in order to get the physically meaningful non-zero-energy edge states,
the amplitude of which must be finite at $n \rightarrow \infty$ in
Eq. (\ref{eq:2-18}), so that we can get the necessary condition for the
edge states $W_{i}=0$ and $W_{\overline{i}}\neq 0$ when $|\lambda_{i}|>1$ and
$|\lambda_{\overline{i}}|< 1$.

Without loss of generality, at first we assume $|\lambda _{1}|>1$ at
some $\{E, k_{y}\}$. In the case, we have $|\lambda _{2}|<1.$
The necessary condition for the existence of edge state is
$W_{1}=0$ and $W_{2}=\eta \neq 0$. From Eq. (\ref{eq:2-19}), the
necessary condition turns to
\begin{equation}\label{eq:2-20}
  \begin{cases}
  A_{11}\Phi_{2,\alpha }+ A_{12}\Phi_{1,\alpha }=0, \\
  A_{21}\Phi_{2,\alpha }+ A_{22}\Phi_{1,\alpha }=\eta(\neq 0).
  \end{cases}
\end{equation}
Where
\begin{eqnarray}\label{eq:2-21}
\begin{cases}
A_{11}=U_{12}^{-1}+GU_{11}^{-1},~~A_{12}=U_{13}^{-1}+(\lambda
+1)U_{11}^{-1},\notag\\
A_{21}=U_{22}^{-1}+GU_{21}^{-1},~~A_{22}= U_{23}^{-1}+(\lambda
+1)U_{21}^{-1},
\end{cases}
\end{eqnarray}
In terms of relations (\ref{eq:2-16}), (\ref{eq:2-17}) and
(\ref{eq:2-12}), we get
\begin{eqnarray}\label{eq:2-22}
A_{11}&=&\frac{C_{12}+GC_{11}}{u_1(a_1-a_2)}=\frac{1}{2\sqrt{a_1^2-4}\sqrt{m}},\nonumber\\
A_{12}&=&\frac{C_{13}+(\lambda+1)C_{11}}{u_1(a_1-a_2)}=\frac{a_2}{2\sqrt{a_1^2-4}\sqrt{m}},\nonumber\\
A_{21}&=&\frac{C_{22}+GC_{21}}{u_2(a_1-a_2)}=\frac{-1}{2\sqrt{a_1^2-4}\sqrt{m}}\nonumber\\
A_{22}&=&=\frac{C_{23}+(\lambda+1)C_{21}}{u_2(a_1-a_2)}=\frac{-a_2}{2\sqrt{a^2_{1}-4}\sqrt{m}}.
\end{eqnarray}
where the relations Eq. (\ref{eq:2-11}), Eq. (\ref{eq:2-12}),
Eq. (\ref{eq:2-16}), and Eq. (\ref{eq:2-17}) have been
used. According to $m\neq 0$ and $a_{1}^2\neq 4$
and $a_{2}^2\neq 4$, $A_{11}$ and $A_{21}$ both
are finite and not zero. If $a_{2}=0$, we have $A_{12}=A_{22}=0$.
The $1st$ Eq. (\ref{eq:2-20}) must lead to
$\Phi_{2,\alpha}=0$ because of $A_{11}\neq 0$, meanwhile the $2nd$
Eq. (\ref{eq:2-20}) showing $A_{22}\Phi_{1,\alpha} = \eta \neq 0$
could not be satisfied. Thus $a_{2}=0$ must result $W_{2}=0$.
Assume all elements $\{A_{ij}\}$ are not zero ($a_2\neq 0$) in our following
discussion. To ensure that both equations in Eq. (\ref{eq:2-20}) should
be satisfied with $\eta \neq 0$, we can write down a non-homogeneous linear
equations with vector $(\Phi_{2,\alpha}, \Phi_{1,\alpha})^{t}$,
therefore Eq. (\ref{eq:2-20}) can be written as
\begin{equation}
    A \left(\begin{array}{c} \Phi_{2,\alpha} \\
            \Phi_{1,\alpha} \end{array}\right)
    = \left(\begin{array}{c} 0 \\ \eta (\neq 0) \end{array} \right)
    = b
\end{equation}
where $A$ is a matrix with the elements $\{A_{i,j}:i,j=1,2\}$, and suppose
$b = ( 0, \eta )^{t}$. The existence of the solution above needs
rank$(A) = $ rank$(A, b)$. After simple but tedious algebraic calculation, we can find that
rank$(A) \neq $ rank$(A, b)$, when $\eta \neq 0$. So we can conclude that
there is no non-zero-energy edge state here. Besides, we can get similar
relations with other $\lambda_{i}>1$ and $\lambda_{\overline{i}}<1$.
The details are neglected here, all the discussion are rigorous. Thus we
rigorously proved that no any edge state exists.

When $m=0$, that leads to $a_{1} = a_{2} = \theta$, $\{\lambda_{i}\}$ are
still the eigenvalues of transfer matrix $T_{0}$. Because the values of
$1st$ and $3rd$ column of $U$ are the same, and those of $2nd$ and $4th$ the
same as well, it is easy to get $\det{U} = 0$, so $U^{-1}$ doesn't exist.
And it is easy to confirm that $ |\lambda_{1}|=|\lambda_{2}|=1 $ and
$|\lambda_{3}|=|\lambda_{4}|=1$, which corresponds to the oscillation of the
wave function, without decay. Meanwhile, under this condition, we can get
$E^{2} = 1-\theta^{2}$. Taking this value back to Eq. (\ref{eq:bulk}),
we can get the relation $2\cos{(k_{x}a/2)} = \cos{(\sqrt{3}k_{y}a/2)}$.
Energies show in Eq. (\ref{eq:bulk}) are complete with both real wave
vectors, so it results that there are just bulk states when $m=0$.

When $a_{1}^{2} = 4$ or $a_{2}^{2} = 4$, $U^{-1}$ does not exist, because
the values of $1st$ and $2nd$ column of $U$ are the same. Without loss
of generality, we can choose $a_{1}^{2} = 4$ in our following discussion, and
it is easy to know $m$ is real, so we have $a_{1}=2$. Then we can easily get
$ |\lambda_{1}|=|\lambda_{2}|=1 $, while at some $k_{y}$, $ |\lambda_{3}|=
|\lambda_{4}|=1$ corresponds to the extended states, and others
$|\lambda_{3}|<1$ and $|\lambda_{4}|>1$ just $\lambda_{3} \cdot \lambda_{4}=1$.
Although $U^{-1}$ doesn't exist, we can suppose to write down it with the
definition of right inverse matrix at $a_{1}\rightarrow 2$, so we can find
at this limitation, the $1st$ and $2nd$ row of $U^{-1}$ are closely the
same in values, while both of them correspond to the extended states. And
besides, we can consider the $3rd$ and $4th$ row which corresponds to the
possible decay wave with a certain $k_{y}$. Then it is easy to get similar
relation as Eq. (\ref{eq:2-19}). In order to get the physically meaningful
edge states, we can get similar relations as Eq. (\ref{eq:2-20}).
Considering about a similar inhomogeneous equations, we can conclude that
there is no non-zero-energy edge state under this condition. Besides, we
can give the similar discussion with $a_{2} = -2$.

In this paper, we give an analytically proof of the non-existence of
the edge states in the semi-infinite AEG. By this method, we show the
bulk energy spectrum and the condition for the existence of the physically
meaningful edge states, according to the rigorous discussion of the
transfer matrix and its eigenstates, we can get the necessary and sufficient
condition for the existence of zero-energy and non-zero-energy edge states,
and finally we find the contradictory condition to show the non-existence
of them. 

\section{Acknowledgments} We should like to thank Dr. X.Z. Yan for helpful
comments. This work is supported by the National Natural Science
Foundation of China under grant  No.10847001 and National Basic
Research Program of China (973 Program) under the grant
(No.2009CB929204, No.2011CB921803) project of China.

\end{document}